# Introduction to UAN Power Equipment Condition Datasets

M. Martin[1], F. Liu[2], J. Sun[3], K.Fiske[4], M.Sagin[5], J.Shackleford[6], N.Hellerstein[7], V.Chong[8]

[1-3] *Auroki Analytics, Vancouver, BC, Canada*

[4-8] *Utility Analytics Network*

Abstract

Power systems are equipment and data intensive. In today's big data era, a large amount of data is being generated from power system equipment via various inspection and testing activities. The industry is encouraged to make optimal decisions for power equipment maintenance, replacement, configuration and planning. Utility companies are trying to leverage the use of big data and advanced analytics for equipment management in order to balance system reliability, performance and cost. To advance such applications and promote collaborations with external researchers, the Utility Analytics Network has made the effort to gather 15 datasets from multiple utility companies in North America on various types of power equipment. These datasets are now shared in public and this paper serves the purpose of providing a detailed introduction to the datasets.

KEY WORDS:  Machine Learning, Data Analytics, Power Equipment, Asset Management,

# 1. Introduction

Big data and machine learning technologies are transforming every industry in today's world. It is doing the same to the power industry which has got more than a hundred years of history. In fact, there are probably more uses cases for machine learning in the power industry due to the existence of a huge amount of data that its massive networks produce continuously, sometimes in real time. Similar to other traditional industries, the value embedded in such data is huge and once released, will be very beneficial to the society, especially in a world where energy conservation and climate change are critical topics that affect the human welfare.

Given such a significance, however, the application of machine learning in this industry have not been researched as well as other traditional industries such as finance, transportation and health care [1-3]. This is probably due to the following reasons:

1. The behaviour and mechanisms of power system are not easy to understand. In contrast, ordinary people are relatively more familiar with finance, transportation and health care industries. This unfamiliarity creates a technical barrier for external researchers such as computer and data scientists to get into the power industry and apply their knowledge and skills to solve the domain problems.

2. The industry has not fully realized the value of its data. Compared to other industries, the power industry is a relatively conservative industry and is more careful with new technologies. Many engineers are not familiar with the concepts of data science and machine learning as well as their value.

However, the above situation has been changing in recent years - more and more power industry datasets have been made available to the public. Examples include smart metering dataset [4], PMU dataset [5], electric vehicle dataset [6], renewable energy generation dataset [7], power line fault dataset [8] and so on.

Our effort focuses on an important aspect of power system business – equipment management. It is well known that the power industry is the most equipment intensive industry. Power system is a giant network that consists of various types of generation, transmission and distribution equipment. This equipment network is so large that it taps to almost every building and every corner in the society that people do not even notice the cables, conductors, transformers and poles behind it – they've taken the access to electricity for granted. The truth is that to make this giant network work as desired, diligent utility engineers and workers are working around the clock to provide this essential service that people seldom see interrupted. One can imagine the daunting task of managing millions of pieces of power equipment, knowing when to inspect, when to replace, how to maintain with respect to owning and maintaining a car which is just one piece of equipment. Manually making decisions for power equipment is very challenging and we believe such decision making can be enhanced and automated by equipment data analytics. To promote further exploration and investigation on power equipment data analytics, the Utility Analytics Network has made the effort to gather 15 datasets from multiple utility companies in North America. This paper intends to provide a detailed introduction to the datasets from a few different angles: data source, purposes, description of each dataset, health indexing method and a machine learning application example to illustrate the use of data.

## 2. Data Source

Sponsored by Auroki Analytics (a Canadian data science consulting company), Utility Analytics Network reached out to 7 utility companies in US and Canada that have adopted Computerized maintenance management system (CMMS) for tracking and storing equipment inspection data. These companies include transmission facility owners, distribution facility owners or a mix of both. To ensure data quality and consistency, when Utility Analytics Network made data requests to the companies, it provided exemplar dataset format. It also did some data processing and organizing work to make sure the formatting among different datasets is consistent to a certain acceptable extent.

Because the inspection data could be intentionally analyzed to infer the reliability performances of utility companies or the product quality of a certain equipment manufacturer, to avoid any potential business liability, the names of the utility companies are omitted. Instead, upon agreement, the service regions of the companies are provided in the dataset descriptions to give users a general idea of the equipment locations.

## 2. Purposes

The datasets mainly intend to serve the following purposes:

1. Help researchers study the condition degradation processes of different types of power equipment;

2. Help researchers develop condition prediction models for different types of power equipment;

3. Help researchers study the relationship between equipment health indexes and inspection conditions;

4. Help researchers study optimal short-term and long-term management strategies for power equipment;

5. Support reliability related computer simulations for distribution and transmission systems.

## 4. Dataset Description

There are in total 15 datasets that have been collected, processed and organized. Their attributes are summarized in Table 1. All the listed equipment is inspected periodically in the utility companies. It is a common practice for utility companies to inspect or maintain their equipment at a certain fixed inspection interval. For different types of equipment, the inspection interval is set differently based on the characteristics of equipment by asset maintenance engineers. The inspection results from different inspection years are also recorded into the CMMS system to help utility engineers trace the history. To reflect this data evolvement, each dataset is organized into a few Excel sheets each of which includes the acquired conditions from one inspection year. The same equipment IDs are shared between different inspection years for cross reference purpose. It should be noted that 2 of the 15 datasets do not have health index included in them and the concept of health index will be explained in Section 5.

Table 1: Dataset Summary

| Equipment Type | Equipment Quantity | Number of Equipment Conditions | Number of Inspection Years Included | Health Index Included? |
|---|---|---|---|---|
| 15 KV XLPE Underground Cable | 2500 | 4 | 4 | Yes |
| 20 KV XLPE Underground Cable | 3943 | 4 | 4 | Yes |
| 138 KV EPR Underground Cable | 4682 | 4 | 4 | Yes |
| 25 KVA Padmount Transformer | 4214 | 4 | 4 | Yes |
| 37.5 KVA Padmount Transformer | 4640 | 4 | 4 | Yes |
| 15 KVA Overhead Transformer | 4467 | 4 | 6 | Yes |
| 25 KVA Overhead Transformer | 4728 | 4 | 6 | Yes |
| Steel Tower | 936 | 2 | 3 | Yes |
| Vaccum-insulated Reclosers | 340 | 5 | 3 | No |
| Oil-insulated Reclosers | 401 | 5 | 3 | No |
| Concrete Pole | 4317 | 2 | 3 | Yes |
| 45-ft Power Pole | 3000 | 7 | 3 | Yes |
| 50-ft Power Pole | 4800 | 7 | 3 | Yes |
| Padmount Switchgear (SF6 Type) | 826 | 5 | 4 | Yes |
| Padmount Switchgear (Air Type) | 1310 | 6 | 4 | Yes |

## 5. Health Indexing Method

Today, many utility companies have adopted the concept of health indexing to quickly determine equipment health statuses. A health index can be scaled from 1 to 5 or other numerical ranges. Equipment with a high health index (such as 5) needs minimum attention and may be recommended for a passive run-to-fail strategy; equipment with a low health index (such as 1) may be recommended for proactive replacement to avoid potential system loss; equipment with a medium health index may be recommended for more frequent inspection and maintenance than the beginning.

In 2016, the international Centre for Energy Advancement through Technological Innovation based in Quebec, Canada (CEATI) surveyed 21 utility companies in US, Canada, Australia and Israel and provided some examples for combining inspection condition attributes to health indexes and also recommended typical actions against different health index levels [9]. In recent years, many utility companies in the world have adopted similar methodology and practices to convert equipment conditions to health indexes for power equipment. Table II showes a typical health indexing example.

Table I: Typical Health Index Definition and Recommendations

| Health Index | Definition | Recommendation |
|---|---|---|
| 5 | No Defect | Regular Monitoring |
| 4 | Minor Defects | Increased Monitoring |
| 3 | Moderate Defects | Scheduled Maintenance/Repair |
| 2 | Significant Defects | Urgent Maintenance/Repair |
| 1 | Serious Defects | Replace |

In comparison with using separate condition attributes to describe a piece of equipment, health index provides a simple indicator for equipment health status and once set up, it can be quickly referenced for decision making.

As listed in Table 1, 13 out of the 15 datasets have health index recorded in the latest inspection while 2 of them do not have health indexes due to a different company practice.

### 6. A Machine Learning Example

To better reveal the usefulness of the dataset, a simple machine learning example is provided. This example intends to study the relationship of different conditions and the health index for the "50-ft Power Pole" dataset. We applied XGBoost classifier to the 2019 inspection record sheet in the dataset. Feature importance is estimated when determining the split node in each boosting tree: for a single boosting tree, feature importance is calculated by the amount that each node improves the performance measure, weighted by the number of observations the node is responsible for. Gini index is used to measure the performance. In the end, feature importance is averaged across all of the boosting trees within the model. Gini index is mathematically given as:

$$Gini = 1 - \sum_{i=1}^{n} p_i^2$$

Where $n$ is the total number of health index classes within a node set and $p_i$ is the probability of health index class $i$ within the node set.

The results are summarized in Figure 1.

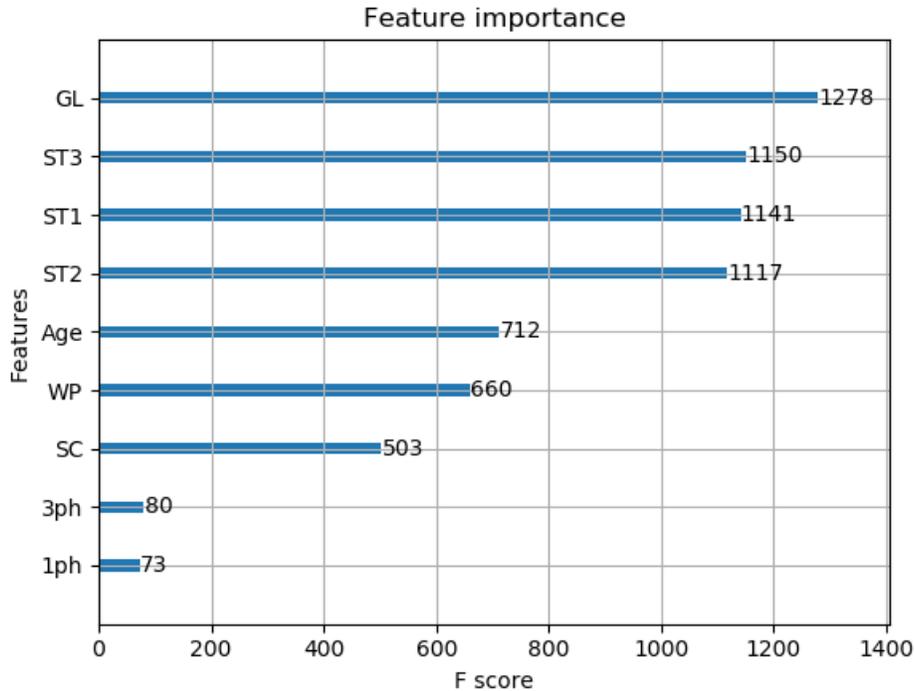

Figure 1: Analysis of the relationship between pole inspection conditions and health index

The importance ranking of pole inspection conditions against its health index are: ground line circumference, shell thickness 3, shell thickness 1, shell thickness 2, age, wood pecker hole, surface condition and the existence of transformer mounted to the pole. This result implies that utility maintenance crews should pay more attention to the ground line circumference and shell thickness when maintaining power poles.

## 7. Conclusions

This paper provides an introduction to the UAN power equipment condition datasets. Data characteristics, structure, purposes and a machine learning application example based on the data are given. The datasets mentioned in Section 4 can be downloaded from Kaggle [10].

## References

[1] Emerson, Sophie, Ruairí Kennedy, Luke O'Shea, and John O'Brien. "Trends and applications of machine learning in quantitative finance." In 8th International Conference on Economics and Finance Research (ICEFR 2019). 2019.